\documentclass{sigchi}

\CopyrightYear{2020}
\setcopyright{acmlicensed}
\doi{https://doi.org/10.1145/3313831.3376725}
\isbn{978-1-4503-6708-0/20/04}
\conferenceinfo{CHI'20,}{April  25--30, 2020, Honolulu, HI, USA}
\acmPrice{\$15.00}

\usepackage{balance}       
\usepackage{graphics}      
\usepackage[T1]{fontenc}   
\usepackage{txfonts}
\usepackage{mathptmx}
\usepackage[pdflang={en-US},pdftex]{hyperref}
\usepackage{color}
\usepackage{booktabs}
\usepackage{textcomp}
\usepackage{float}

\usepackage{microtype}        
\usepackage{ccicons}          

\usepackage{todonotes}

\def\plaintitle{An Intermittent Click Planning Model}

\def\emptyauthor{}
\def\plainkeywords{Click model; pointing; intermittent control; cue integration; temporal pointing; internal clock.}

\makeatletter
\def\url@leostyle{%
  \@ifundefined{selectfont}{
    \def\UrlFont{\sf}
  }{
    \def\UrlFont{\small\bf\ttfamily}
  }}
\makeatother
\def\pprw{8.5in}
\def\pprh{11in}

\definecolor{linkColor}{RGB}{6,125,233}
\hypersetup{%
  pdftitle={\plaintitle},
  pdfauthor={\emptyauthor},
  pdfkeywords={\plainkeywords},
  pdfdisplaydoctitle=true, 
  bookmarksnumbered,
  pdfstartview={FitH},
  colorlinks,
  citecolor=black,
  filecolor=black,
  linkcolor=black,
  urlcolor=linkColor,
  breaklinks=true,
  hypertexnames=false
}
\usepackage{mathtools, nccmath}

\usepackage[ruled]{algorithm2e} 

\usepackage[noend]{algpseudocode}

\newcommand{\me}{\mathrm{e}}

\definecolor{bcolor}{rgb}     {1.0,0.0,0.0}
\definecolor{ecolor}{rgb}     {0.0,0.0,1.0}

\newcommand\blfootnote[1]{%
  \begingroup
  \renewcommand\thefootnote{}\footnote{#1}%
  \addtocounter{footnote}{-1}%
  \endgroup
}

\newcommand{\finalAdd}[1]{#1}
\newcommand{\finalDel}[1]{}

\usepackage{enumitem}

\setitemize{noitemsep,topsep=0pt,parsep=0pt,partopsep=0pt}

\begin{document}

\title{\plaintitle}
\numberofauthors{1}
\author{
  \alignauthor{Eunji Park and Byungjoo Lee$^*$\\
    \affaddr{KAIST, Republic of Korea}\\    
    \email{\{eunji.park,byungjoo.lee\}@kaist.ac.kr}}\\
}

\maketitle

\begin{abstract}
\finalDel{Pointing is the most basic task of human-computer interaction (HCI) in which a user tracks a target with a pointer and must perform a click action when the pointer is positioned within the target.}
\finalAdd{Pointing is the task of tracking a target with a pointer and confirming the target selection through a click action when the pointer is positioned within the target.}
Little is known about the mechanism by which users plan and execute the click action in the middle of the target tracking process.
The Intermittent Click Planning model proposed in this study describes the process by which users plan and execute optimal click actions, from which the model predicts the pointing error rates. 
In two studies in which users pointed to a stationary target and a moving target, the model proved to accurately predict the pointing error rates
\finalAdd{($R^{2}=0.992$ and $0.985$, respectively).}
The \finalDel{ICP }model has also successfully identified differences in cognitive characteristics among first-person shooter game players.
\end{abstract}




\section{Introduction}
\finalDel{During the pointing process, a user cannot obtain a target simply by moving the pointer toward the target.
The user must perform a click action to press the mouse button, at the appropriate moment when the user perceives that the pointer is positioned within the target.}
\finalAdd{When pointing through an input device such as a mouse or trackpad, users typically perform a click action to confirm the target selection. At that time, for successful target selection, the mouse button must be pressed at the appropriate moment when the pointer is located within the target.}
Here, several questions arise: 1) when and how do users plan and execute a click action during pointing? 2) what factors influence the accuracy and precision of the click action? This study explores the user's \emph{click planning and execution process}, which has rarely been considered in previous pointing studies in human-computer interaction (HCI).

Pointing is a task in which the tracking and the click action are combined into one, and both actions can be blended in different ways in time and space.
For example, a careful user can plan and perform a click action after it is certain that the pointer is positioned within the target.
On the other hand, users who are willing to take a risk can \emph{anticipate} when the pointer will be positioned within the target and begin planning and executing the click action before the pointer actually enters the target.

To the best of our knowledge, extant research has not explored the process of planning and executing click in depth.
For example, Fitts' law \cite{fitts1954information,mackenzie1992fitts}, a key model for user's pointing performance, yields an accurate prediction of the overall pointing completion time. However, there is still room for providing assumptions and explanations for the mechanism of the user clicking on the target. 
\finalDel{Besides, the control theoretic models of pointing \cite{muller2017control,bye2008bump, aranovskiy2016modeling, aranovskiy2019switched} effectively predict the user's tracking movement to the target, but they do not include the detailed mechanism of the click planning and execution process in model construction and simulation.}
\finalAdd{Control theoretic models \cite{muller2017control,bye2008bump, aranovskiy2016modeling, aranovskiy2019switched}, on the other hand, successfully explain the fundamental principles of human motor planning and execution, but only limited explanations have been given for click action so far compared to tracking movement in pointing.}
\blfootnote{\small*Corresponding author\normalsize}

\subsection{Effect of Click Plan Quality on Pointing Performance}
Since a click action is typically performed at very short distances and duration (less than 130 ms) \cite{zhai2002performance, mackenzie1999design}, it may have a small impact on the overall task completion time.
However, since a pointing endpoint is ultimately determined from a click action, the quality of the click plan can affect the end point distribution or error rate ($ER$), another key performance metric of pointing.
The quality of a click here means how accurate and precise a click action can be planned and executed in \emph{time}, as the click action itself usually does not have to be spatially accurate or precise; usually, a finger is already placed on the click button, and clicking requires a very short travel distance. In other words, the Fitts' index of difficulty (ID) of the click action is close to zero.
\begin{figure}[t]
	\centering
    \includegraphics[width=0.95\columnwidth]{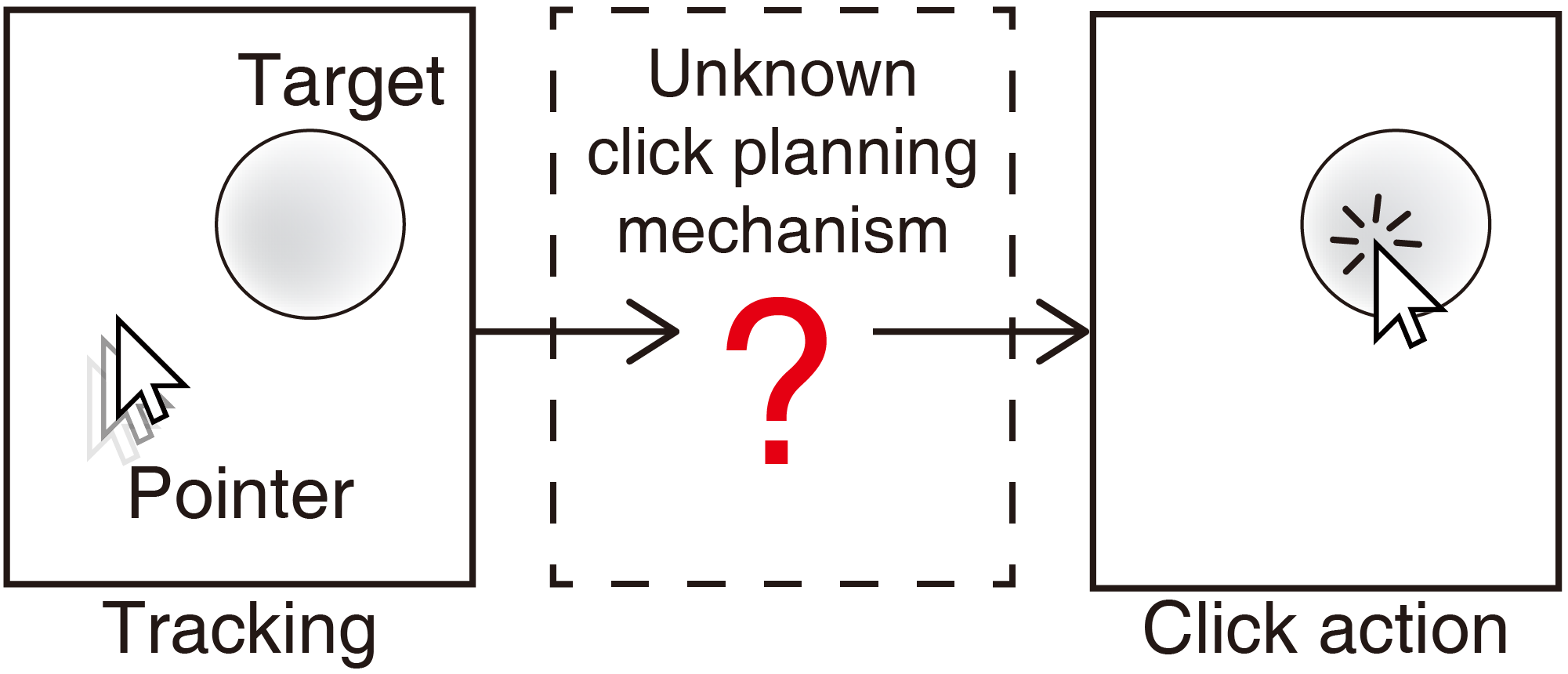}
    	\caption{This study present a computational model for predicting the pointing error rates based on an understanding of the cognitive mechanism in which click action is planned and executed by the user.
    }
  ~\label{fig:question}
~\vspace{-8mm}
\end{figure}
If users click in a situation where they can easily estimate \emph{when} their pointer will be positioned within the target, the quality of the click plan will be high, and the variance of the endpoint distribution and the pointing error rate will be low.
However, when users want to perform a quicker pointing, they try to click in a situation where it is difficult to estimate \emph{when} the pointer be located in the target, the quality of the click plan will be low, and the end point variance and pointing error rate will have a higher value.
This study ultimately aims to present a model for predicting the pointing error rates based on an understanding of the cognitive mechanism in which click timing is estimated by the user.

\subsection{Overview of Intermittent Click Planning Model}

This study presents a cognitive model called the intermittent click planning (ICP) model, which is a predictive model of pointing error rates that includes explicit consideration of the click planning process.

The ICP model essentially regards the click planning process as a \emph{timing decision problem} given to the user's internal clock.
The ICP model describes the cognitive process in which the user's internal clock estimates the optimal timing of click actions while the pointer approaches the target.
To this end, the model considers \finalAdd{six predictor variables} (target position $\vec{p}_{t}$, cursor position $\vec{p}_{c}$, target velocity $\vec{v}_{t}$, cursor velocity $\vec{v}_{c}$, target size $W$, period of click repetition $P$) and \finalAdd{four free parameters} ($c_\sigma$, $c_\mu$, $\nu$, $\delta$) that determine the quality of the user's click planning process by referring to previous studies of sensorimotor synchronization \cite{wing1973response, repp2005sensorimotor, lee2016modelling} and human reaction \cite{ratcliff1978theory, lee2018moving}.
\finalAdd{Predictor variables} describe the situation where the user's pointer is tracking the target, such as the velocity of the pointer and the target, or the target size \finalAdd{(i.e., similar to $D$ and $W$ variables in Fitts' law)}.
\finalAdd{Free parameters of the model} are constants representing the cognitive characteristics of users, for example, the precision of the internal clock and the drift rate of visual encoding \finalAdd{(i.e., similar to $a$ and $b$ parameters in Fitts' law)}.
Once these ten values have been determined, the ICP model can predict the pointing error rates ($ER$).

The ICP model is based on two important assumptions about users.
First, the model assumes that the user is an \emph{intermittent controller}.
In studies of human aimed movement, intermittent control refers to the process by which humans discontinuously control the parameters that produce continuous motion \cite{bye2008bump, neilson1988internal, craik1947theory}.
When analyzing the pointing trajectories, the user's tracking movement is comprised of several submovements, which manifests that intermittent control is taking place.
Due to such intermittency, the model assumes that users build their click plan during the last submovement that was executed just prior to the click.
Otherwise, if click actions were planned earlier than that, the plan would become obsolete because intermittent updates of tracking movements later make the tracking situation different from the situation where the click was planned.
With this assumption, the model only needs to consider the last submovement just prior to the click, not the whole tracking trajectory when predicting the error rates.

Second, ICP model assumes that the user is a statistical encoder that makes optimal use of the externally provided information that allows for estimation of click timings.
Based on \emph{cue integration theory} \cite{ernst2002humans}, the model shows how users integrate click timings estimates from multiple sensory channels into a single click timing estimate, through the maximum likelihood estimation (MLE) process.

We conducted two user studies to verify the explanatory power of the model.
The first study deals with pointing at a stationary target, and the second study concerns pointing at a moving target with constant velocity.
In both studies, our model proved to accurately predict the user's pointing error rates ($R^{2}=0.992$ and $0.985$, respectively).
We also observed that the user's internal factors obtained by fitting the model to the data were \emph{kept at the same level regardless of whether the target was moving or not}.
This shows that our model successfully represents the actual cognitive process, not simply overfitting the data with a number of free parameters.

Predicting error rates in pointing has been of great interest for many reasons, from everyday interactions like touch \cite{bi2013ffitts, bi2013bayesian, bi2016predicting} and typing \cite{weir2012user,yi2015atk,zhu2018typing, findlater2011typing, findlater2012personalized, gunawardana2010usability} to more dynamic interactions like games and music \cite{lee2019geometrically, lee2016modelling, lee2018moving}. 
In general, error rate models has been used for evaluating the performance of a given interface, designing an appropriate level of difficulty for a game, or profiling different user populations.

To the best of our knowledge, this is the first model that describes the mechanism by which users plan and execute a click action during pointing.
Our model does not conflict with existing pointing models, such as Fitts' law or control theoretical models; rather, it provides room for refinement of existing models with a thorough consideration of the click process.

\section{Related Work}
\subsection{Studies on Click Planning and Execution Process}
A click is an input action in which a button is pressed on a keyboard or mouse to carry out a computer operation.
The click is the most basic task of HCI, but it has been studied in the past as a secondary task required to confirm the target selection in pointing. 
In this respect, a click is considered to be performed sequentially based on minimal cognitive effort as long as the pointer is located within the target.
For example, early GOMS family models measured total pointing time as a simple sum of pointer movement time and click action duration \cite{kieras2001using,card1980keystroke}.
This is in line with the fact that in studies of the Fitts' law, a click action has been regarded as a non-information component that simply adds to the trial completion time without affecting the throughput of the pointing process \cite{zhai2004characterizing}.

However, the pointing time predicted by the GOMS models is longer than the actual pointing time \cite{hornof1999cognitive, john2004predictive}. This problem comes from the assumption that the click action is performed sequentially after tracking movement. Therefore, later GOMS models assumed the user's anticipatory click action \cite{john2004predictive}. The models assume that the click action can be planned in advance while tracking the target and can also be performed in a chunk with the tracking movement \cite{card1980keystroke}. This is similar to the perspective of the ICP model in this study. However, the existing GOMS models differ from the ICP model. They do not describe the cognitive processes by which users plan and perform click actions, and also assume that perfect users perform error-free clicks \cite{card1980keystroke, holleis2007keystroke}.

\subsection{Temporal Pointing Studies}
Recently, several studies have modeled the user's performance in a special task where there is no physical movement other than click action (i.e., pure clicking task). In that task, called \emph{temporal pointing} \cite{lee2016modelling, lee2018moving,lee2019geometrically}, the user must acquire a target that is only selectable for a short time, such as a blinking target or a target moving to a selection region, by activating a click event. To successfully accomplish this task, the user must activate the click at an accurate and precise timing by estimating \emph{when} the target will be selectable, using his or her own internal clock. 

In the temporal pointing process, two external cues are given that allow the user to estimate the optimal click timing \cite{lee2018moving}: (1) a temporal repetition cue and (2) a visually perceivable movement cue.
The temporal pointing models describe how the user encodes those external temporal cues to estimate when the target will be selectable, and consequently predict the user's click error rates.
Based on the click planning mechanism described by the temporal pointing model, by adding additional assumptions about the user's intermittent control process, the ICP model in this study describes how users plan and perform click actions during pointing.

\subsection{Predicting Error Rates in Stationary Target Pointing}
In the pointing task to a stationary target, the user's pointing error rates change with the time constraints given to the user. In general, the shorter the time limit given to the user, the higher the error rates. This is called the speed-accuracy trade-off \cite{plamondon1997speed} phenomenon. Based on empirical findings on this phenomenon, there are various models that predict the pointing error rates. The present study does not cover all of the models in detail, and we refer interested readers to \cite{wobbrock2008error}.
In this section, we introduce a state-of-the-art model among them.

Wobbrock et al. \cite{wobbrock2008error,wobbrock2011modeling} derived a model that predicts the pointing error rates of users in pointing when time pressure varies.
Their model is derived from Fitts' law, and the error rate prediction can be done through a closed form equation:
\begin{equation}
\small
ER=1-erf\Bigg\{\frac{1}{D\sqrt{2}}\left[2.066\cdot W(2^{\frac{MT_{e}-a}{b}}-1)\right]\Bigg\}
\label{eq:wobbrockmodel}
\end{equation}
where $a$ and $b$ are free parameters inherited from Fitts' law, $D$ is the target distance, and $W$ is the width or size of the target.
$1/b$ has been called index of performance or throughput.
$MT_{e}$ is the movement time \emph{actually} taken in pointing and is an independent variable of the model.
The model describes the user's error rate successfully for both one-dimensional \cite{wobbrock2008error} and two-dimensional \cite{wobbrock2011modeling} pointing but essentially cannot hold in situations where Fitts' law is violated \cite{chapuis2011effects, adam2006moving, chapuis2008small, glazebrook2015one, wallace1983visual, zhai2004speed}.
The model also does not account for the user's error rate in pointing to a moving target.
Wobbrock's model is used as the baseline for the first user study in the present paper, which deals with pointing to a stationary target.

\subsection{Predicting Error Rates in Moving Target Pointing}
Pointing research has been conducted mainly on stationary targets. However, pointing to moving targets is often done in applications such as games and music. Moving target pointing has been studied in cognitive psychology, called moving-target interception  \cite{belisle1963accuracy} or anticipation-coincidence \cite{fajen2007behavioral} studies. However, in the field of HCI, research on moving target pointing has started only recently \cite{huang2018understanding, lee2018moving, lee2016modelling, casallas2015prediction, casallas2013towards, al2011moving}. 

Unlike a stationary target, in a pointing process on a moving target, the positional relationship between the target and the pointer constantly changes even if the pointer does not move. Therefore, it is difficult to model the pointing situation with a few predictor variables such as the target distance and target width. For example, four or more different pointing strategies (pursuit, head-on, receding, and perpendicular) can be characterized \cite{tresilian2005hitting}. This makes it difficult to build a model that predicts the error rates in a moving target pointing task. 

Recently, Huang and his colleagues \cite{huang2018understanding, huang2019modeling} have suggested a model called Ternary-Gaussian that predicts endpoint distribution from the given target velocity ($V$) and target width ($W$) condition. Based on the assumption that the distance to the target does not affect the user's pointing endpoint distribution, the model can neglect the relative positional relationship between the moving target and the pointer. The model predicts the pointing endpoint distribution as a 2D Gaussian distribution:
\begin{equation}
\small
\begin{split}
    \sigma_{t}&=\sqrt{d_{t}+e_{t}V^{2}+f_{t}W^2+g_{t}V/W}\\
    \sigma_{n}&=\sqrt{d_{n}+e_{n}V^{2}+f_{n}W^{2}}\\
    \mu_{t}&=a_{t}+b_{t}V+c_{t}W \;\;\;\text{and}\;\;\; \mu_{n}=0
    \end{split}
    \label{eq:huangmodel}
\end{equation}
where $\mu_t$ and $\sigma_t$ are the mean and standard deviation of the Gaussian distribution in the target velocity direction, respectively, and $\mu_n$ and $\sigma_n$ are the mean and standard deviation of the Gaussian distribution in the normal direction to the target velocity, respectively. $a$, $b$, $c$, $d$, $e$, and $f$ are ten free parameters of the model that must be determined from the user experiments. The pointing error rate can be obtained by integrating the Gaussian distribution outside the target area.
The Ternary-Gaussian model is used as the baseline for the second user study in this paper, which deals with pointing to a moving target.

\section{Intermittent Click Planning Model}
\subsection{Task and Problem Formulation}
The ICP model concerns the general 2D pointing task. The task involves a user moving a pointer over a target and activating a click event. If a click event occurs when the pointer is within the target, the trial is considered successful; if it occurs outside the target, the trial is deemed failed. When the user repeats the pointing, the ratio of failed trials to total number of trials is the pointing error rate ($ER$).

The ICP model assumes that the target tracking movement and the click action are each \emph{separately} planned and executed by the user. Among them, the ICP model concentrates on the modeling of the click input movement. 
More specifically, in the ICP model, it is assumed that the click action is planned and executed in the middle of the target tracking movement.
Thus, how the target tracking movement of the pointer is being performed will affect the quality of the planned click action. The situation where the pointer is tracking the target can be characterized by the following variables:

\begin{itemize}
\small
    \item Target velocity ($\vec{v}_{t}$): The two-dimensional vector representing the velocity at which the target moves
    \item Target position ($\vec{p}_{t}$): The two-dimensional vector representing the location of the target
    \item Target size ($W$): The width of the target (diameter if the target is circular)
    \item Pointer velocity ($\vec{v}_{p}$): The two-dimensional vector representing the velocity at which the pointer moves
    \item Pointer position ($\vec{p}_{p}$): The two-dimensional vector representing the location of the pointer
    \item Mean period of click repetition ($P$): When clicks are repeated this is the average of the preceding inter-click time intervals (or average click period).
\end{itemize}
Note that the above variables can also explain the pointing situation for a stationary target (i.e., $\vec{v}_{t}=0$).

\subsection{Model Derivation}
\subsubsection{Two-level Process of Click Planning and Execution}
\begin{figure}[t] 
	\centering
    \includegraphics[width=0.9\columnwidth]{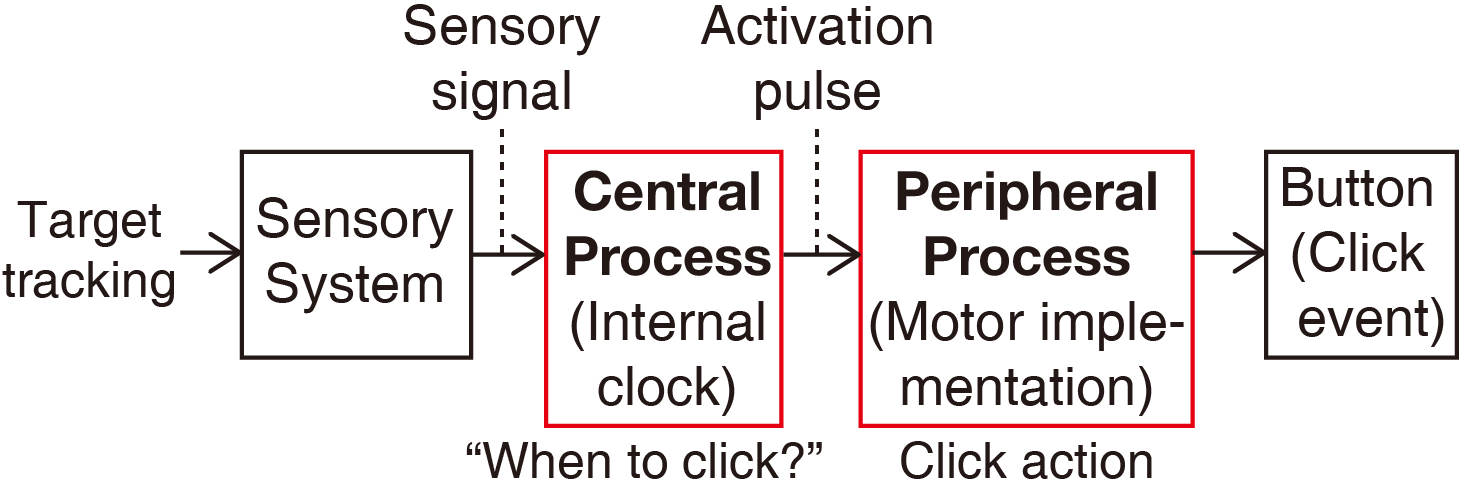}
    	\caption{The ICP model consists of two-level process: The \emph{central process} encodes a given sensory signal, producing activation pulses that trigger the click action. The \emph{peripheral process} executes the click action as soon as the internal clock generates an activation pulse.}
  ~\label{fig:twoprocess}
   ~\vspace{-5mm}
\end{figure}

The ICP model assumes that the planning and execution of click action is based on a two-level process that contains a \emph{central} and a \emph{peripheral} level (see Figure \ref{fig:twoprocess}). The central process is responsible for estimating the optimal click timing from a given target tracking situation. The peripheral process serves to actually implement the click action at the timing estimated by the central process. 

The central process is performed by the user's internal clock. In a given tracking situation, the user's internal clock is given several sensory signals that allow for estimation of optimal click timing. The internal clock then encodes the sensory signals to estimate the optimal click timing, and generates a trigger pulse to activate the click action at the estimated optimal click timing. This trigger pulse is sampled from a probabilistic distribution and has a specific mean and variance because a given sensory signal is not completely reliable.

The peripheral process is performed by the user's motor system. When the internal clock generates an activation pulse, the peripheral process executes a predefined click action as soon as possible. The click action here refers to the very short movement of the user's finger pressing the button. However, there is an inherent delay in the motor system. As a result, the user's click action is always slightly later than the internal clock's intended timing. This delay is also a stochastic value with a specific mean and standard deviation.
According to earlier models of sensorimotor synchronization \cite{wing1973response, vorberg1996modeling}, the trigger pulse and the motor delay can be assumed to be probabilistic independent.

As a result, the quality of the user's click actions depends on the following factors in this model: (1) the quality of the trigger pulses generated by the internal clock and (2) the quality of the motor delay. For the construction of a simple and effective model, we assume that the motor process's quality is much higher (mean and variance close to zero) than that of the internal clock, so that the effect of the motor delay on the click action can be ignored. This is a plausible assumption because click action has a Fitts' $ID$ that is negligible.

\subsubsection{Intermittent Click Planning}
It is known that a person is an intermittent controller when performing target tracking movements. Intermittent control refers to the process of intermittently modifying the original motor plan while humans are performing their movements. That is, the new motor plan intermittently replaces the existing motor plan, and the human prepares the next motor plan while the current motor plan is executed \cite{neilson1988internal, bye2008bump, bye2010bump}. As a result of intermittent control, the trajectory of a person's aimed movement shows inherent discontinuities and is comprised of several submovements. 
Because of such intermittent updates of the motor plan, it is not correct to assume that planning of a click action takes place during the entire target tracking movement. Rather, we assume that the plan for the click action is made during the execution of the last submovement just before the click. If the click action is planned before that, it will be meaningless due to the subsequent intermittent control process. 

The start and end of a submovement can be found by tracking the local maximum and minimum from the speed profile of the trajectory \cite{lee2016autogain}.
Assuming that the time at which the last submovement started is $t_{sub}$, all model derivations after this section deal only with the pointing situation since that point. 

\subsubsection{Implicit Aim Point in Time}
In order for the click action to be attempted at the appropriate timing, the user's internal clock must first estimate the timing of two major events that occur during the target tracking movement. The first event is the moment the user's pointer first contacts the target ($t=t_{enter}$). The second event is the moment the user's pointer exits the target ($t=t_{exit}$). The user must activate the click action between these two events to make the target acquisition successful.

\begin{figure}[t] 
	\centering
    \includegraphics[width=0.6\columnwidth]{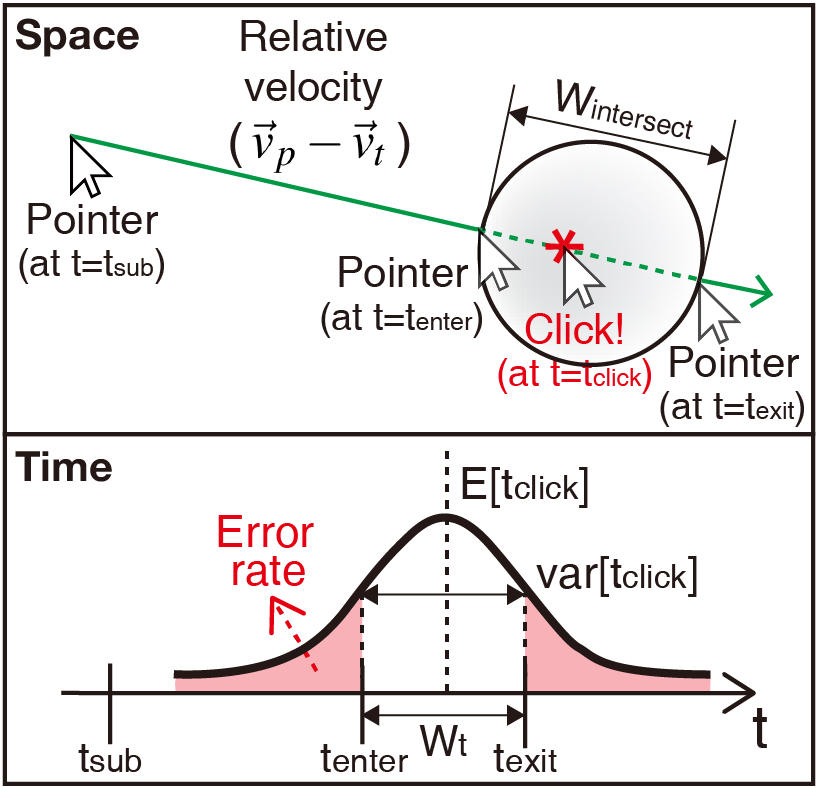}
    	\caption{The ICP model describes the cognitive process in which a user estimates the optimal click timing ($t_{click}$) in the last submovement of the target tracking process.}
  ~\label{fig:winterect}
   ~\vspace{-5mm}
\end{figure}

Let $W_t$ be the time interval between two events. $W_t$ is the duration from the moment the pointer first touches the target until it passes completely through the target: \finalDel{Thus, from the relative velocity vector between the target and the pointer ($\vec{v}_{p}-\vec{v}_{t}$), it can be calculated as follows:}
\begin{equation}
    W_{t}=W_{intersect}/||\vec{v}_{p}-\vec{v}_{t}||
\end{equation}
where $\vec{v}_p$ is the velocity vector of the pointer and $\vec{v}_t$ is the velocity vector of the target. $W_{intersect}$ is the length of the line where the movement trajectory of the pointer intersects with the target (see Figure \ref{fig:winterect}).
\finalAdd{If the pointer does not penetrate the target during the last submovement, the value of $W_{intersect}$ is simply zero.}

The ICP model assumes that the user can encode the $t_{enter}$ and $t_{exit}$ from the sensory signals given in the pointing situation and can determine the appropriate click timing ($t_{click}$) within $W_{t}$. The click timing is expressed relative to $W_{t}$ as follows:
\begin{equation}
    t_{click}= c_{\mu} \cdot W_{t}
\end{equation}
In the above equation, time is defined with $t_{enter}$ as $0$; $c_{\mu}$ is called the user's \emph{implicit aim point}. If $c_{\mu}$ is $0$, $t_{click}= t_{enter}$; if $c_{\mu}$ is $1$, $t_{click}$ = $t_{exit}$.

\subsubsection{Input Distribution in Time}
Since negligible motor delay is assumed, a user's click timing distribution is determined solely from the distribution of click timings estimated by the internal clock.
In this case, all cases where $t_{click}$ is earlier than $t_{enter}$ or later than $t_{exit}$ are failed inputs.
In sensorimotor synchronization studies \cite{repp2005sensorimotor}, a person's response timing distribution can generally be regarded as Gaussian $\mathcal{N}(\mu,\sigma^{2})$, and its mean ($\mu$) and variance ($\sigma^{2}$) can be expressed as (see Figure \ref{fig:winterect} bottom):
\begin{equation}
    \mu=\text{E}[t_{click}]\;\;\;\text{and}\;\;\;\sigma^2=\text{Var}[t_{click}]
\end{equation}

\subsubsection{Modeling Mean of Click Timing Estimates}
The ICP model assumes a user's \emph{constant implicit aim point} ($c_{\mu}$) to account for the mean of the click timing distribution. In other words, the mean of the user's click timing distribution is presumed to be located at a constant ratio of the length of $W_{t}$. For example, if the user's implicit aim point is 0.25, the mean value $\mu$ of the click distribution would be 25 ms for $W_t$ 100 ms, and 50 ms for $W_t$ 200 ms (from $t_{enter}=0$). The constant implicit aim point of the user has been demonstrated experimentally from previous temporal pointing studies \cite{lee2016modelling,lee2018moving,lee2019geometrically}. Based on this assumption, the mean ($\mu$) of the input distribution can be expressed as:
\begin{equation}
    \mu=\text{E}[t_{click}]=\text{E}[c_{\mu} W_{t}]=c_{\mu} W_{t}
    \label{eq:mu}
\end{equation}

\subsubsection{Modeling Variance of Click Timing Estimates}
While approaching the target, users can estimate click timing from two different sensory cues. The first cue is a \emph{temporal structure cue}; if the click has been performed at regular intervals or at a specific rhythm, the user can also estimate the timing of the next click without additional information \cite{lee2016modelling}.
The second is a \emph{visually perceivable movement cue} that is encoded from the relative position and velocity between the pointer and the target. Knowing the distance between the pointer and the target and the relative velocity at which the pointer approaches the target, users can estimate the timing when the click action should be performed \cite{lee2018moving, tresilian1991empirical}.

The click timings estimated from each cue have different standard deviations. The standard deviation of the click timing estimated from the temporal structure cue ($\sigma_{t}$) is known to increase in proportion to the period of repetition ($P$). This is known as the scalar property of the internal clock \cite{repp2005sensorimotor, gibbon1984scalar}. For example, clapping every 5 seconds will have a higher timing variability than clapping once every second:
\begin{equation}
    \sigma_{t} \propto P
    \label{eq:sigmat}
\end{equation}

Next, the standard deviation of the click timing encoded from the visual movement cue ($\sigma_{v}$) can be modeled as a function of the time interval at which the user can observe the relative movement between the pointer and the target (i.e., cue viewing time $t_{c}$):
\begin{equation}
    \sigma_{v}\propto  1/(\me^{\nu t_{c}}-1)+\delta
    \label{eq:sigmav}
\end{equation}
This equation provides an intuitive understanding of the reliability of the click timing encoded from the visual cue \cite{lee2018moving}. If cue viewing time is sufficiently long ($t_{c}\rightarrow \infty$), $\sigma_{v}$ converges toward $\delta$. $\delta$ represents the minimum standard deviation of the click timing the user can encode from the visual cue. On the other hand, if cue viewing time is very short ($t_{c}\rightarrow 0$) this is the same situation as when no visual cue exists, where $\sigma_{v}$ diverges to infinity. $\nu$ represents the rate at which the user encodes click timing from the visual cue. The higher this is, the more precise click timing can be planned for the same $t_{c}$. $\nu$ is also called the drift rate in models of human reaction \cite{ratcliff1978theory}.

Since we consider only the last submovement in the click model due to the user's intermittent control process, it can be assumed that the cue viewing time ($t_{c}$) is the time from the start of the submovement ($t_{sub}$) to the time of the click ($t_{click}$):
\begin{equation}
    t_{c}=t_{click}-t_{sub}
    \label{eq:tc}
\end{equation}

Finally, by defining $c_{\sigma}$ as the parameter that determines the proportionality of Equations \ref{eq:sigmat} and \ref{eq:sigmav}, $\sigma_{t}$ and $\sigma_{v}$ can be expressed as follows:
\begin{equation}
    \sigma_{t} = c_{\sigma} P \;\;\text{and}\;\;
    \sigma_{v} = c_{\sigma} (1/(\me^{\nu t_{c}}-1)+\delta)
\label{eq:variances}
\end{equation}

\subsubsection{Integration of Click Timing Estimates}

\begin{figure}[t] 
	\centering
    \includegraphics[width=1\columnwidth]{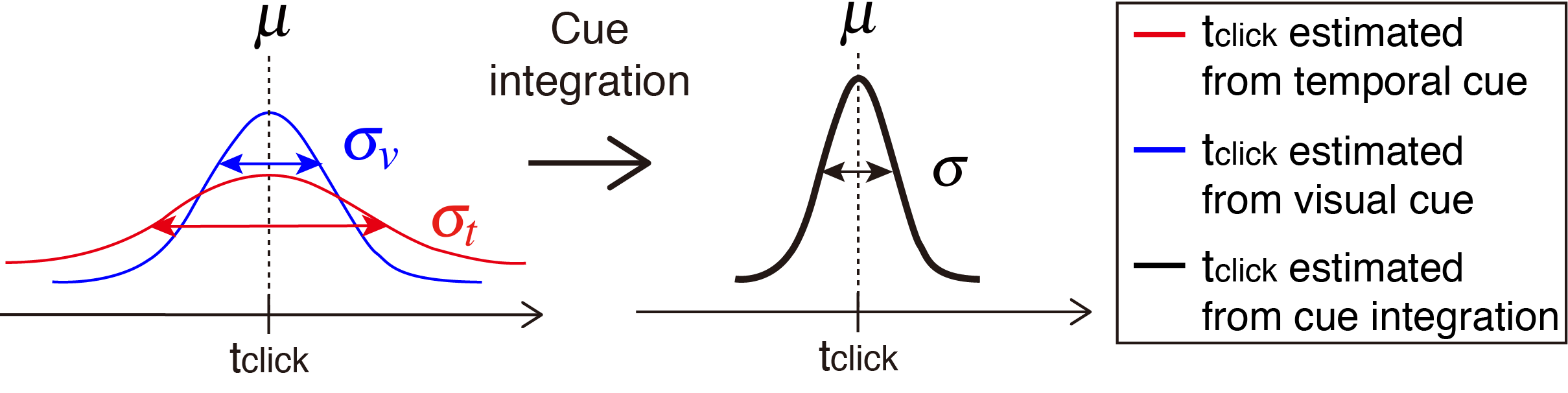}
    	\caption{The final click timing is encoded separately from each of the temporal structure cue and the visual cue.}
  ~\label{fig:cues}
   ~\vspace{-6mm}
\end{figure}

Cue integration theory \cite{ernst2002humans, lee2018moving,oulasvirta2018neuromechanics} describes the process by which humans integrate multiple estimates of a physical quantity into a single estimate. According to the theory, humans encode the given information in a statistically optimal way, which can be expressed by the maximum likelihood estimation (MLE) process. For example, let's say a person wants to estimate the size of a ball. The person can estimate the size of the ball separately through visual and haptic sensations. If the estimates from each sensory channel have different variance, $\sigma_{1}^2$ and $\sigma_{2}^2$, then the variance of the person's final estimate of ball size is expressed by MLE as:
\begin{equation}
    \sigma^{2}=\sigma^{2}_{1}\sigma^{2}_{2}/(\sigma^{2}_{1}+\sigma^{2}_{2})
\end{equation}
Through this process, the reliability of the integrated estimate is higher than the reliability of individual estimates.

Based on the cue integration theory, we can also model the variance of the final click timing that the user estimates during the target tracking process.
As described earlier, click timing can be estimated separately from two different sensory cues, and the variance of those estimates can be expressed as $\sigma_{t}^2$ and $\sigma_{v}^2$ in Equation \ref{eq:variances}.
When those two estimates of click timings are integrated into a single click timing by MLE, the variance of the final click timing estimate ($\sigma^{2}$) can be expressed as:
\begin{equation}
    \sigma^{2}=\frac{\sigma^{2}_{t}\sigma^{2}_{v}}{(\sigma^{2}_{t}+\sigma^{2}_{v})}=  \frac{c_{\sigma}^{2} \cdot P^{2}}{(1+(P/(1/(\me^{\nu t_{c}}-1)+\delta))^{2})}
    \label{eq:sigma}
\end{equation}
\subsubsection{End Point Distribution and Pointing Error Rate}
Finally, the user's click timing distribution is a 1D Gaussian distribution $\mathcal{N}(\mu,\sigma^{2})$ with mean $\mu$ (Equation \ref{eq:mu}) and variance $\sigma^2$ (Equation \ref{eq:sigma}) on the time axis. In this distribution, cases where the click input timing $t_{click}$ is between 0 and $W_t$ are successful in acquiring the target, and cases where the click input timing $t_{click}$ is smaller than 0 or greater than $W_t$ are failing in the target acquisition. So by subtracting the integral of the input distribution from 0 to $W_{t}$ from 1.0, we can get the expected pointing error rate ($ER$):
\begin{equation}
\begin{split}
ER &= 1-\int_{0}^{W_{t}-\mu} \mathcal{N}(t)dt  - \int_{0}^{\mu} \mathcal{N}(t)dt\\
	&= 1-\frac{1}{2}\left[erf(\frac{W_{t}-\mu}{\sigma\sqrt{2}}) 	+erf(\frac{\mu}{\sigma\sqrt{2}})\right]
\end{split}
\label{eq:er}
\end{equation}
\vspace{-5mm}
\finalAdd{
\begin{equation*}
	= 1-\frac{1}{2}\left[erf(\frac{(1-c_{\mu})W_{t}}{\sqrt{{\frac{2c_{\sigma}^{2} \cdot P^{2}}{1+(\frac{P}{\frac{1}{\me^{\nu t_{c}}-1}+\delta})^{2}}}}}) + erf(\frac{c_{\mu}W_{t}}{\sqrt{\frac{2c_{\sigma}^{2} \cdot P^{2}}{1+(\frac{P}{\frac{1}{\me^{\nu t_{c}}-1}+\delta})^{2}}}})\right]
\end{equation*}
}

\begin{table}
\centering
\caption{Comparison between baseline models and ICP model}
\label{tablecompare}
\scriptsize
\begin{tabular}{|p{1.7cm}|p{1.4cm}|p{0.9cm}|p{1.0cm}|p{1.5cm}|}
\hline
&No. free parameters&Moving Target & Stationary Target& Physical meaning of free parameters\\ \hline
Baseline 1 \cite{wobbrock2008error}&2&No&Yes&Yes \\ \hline
Baseline 2 \cite{huang2019modeling}&10&Yes&No&No \\ \hline
ICP model&4&Yes&Yes&Yes \\ \hline
\end{tabular}
\vspace{-4mm}
\end{table}
\subsection{Summary of Free Parameters in the ICP Model}
The ICP model has \emph{four} free parameters ($c_{\sigma}$, $c_{\mu}$, $\nu$, $\delta$) that must be calibrated through user experiments, such as the slope and $y$-intercept in Fitts' law. The larger the number of free parameters, the better the model can describe the data, but at the same time, overfitting problems can arise. The baseline model of stationary target pointing \cite{wobbrock2008error} has two free parameters, and the baseline model of moving target pointing \cite{huang2019modeling} has ten free parameters. With just four parameters, the ICP model can effectively predict the pointing error rate for both moving and stationary targets (see Table \ref{tablecompare}). Also, each parameter of the ICP model has a clear cognitive meaning.
The meaning of each free parameter of the ICP model is:

\begin{itemize}
\item  $c_{\sigma}$ represents the encoding precision of the user's internal clock. The higher the $c_\sigma$, the lower the performance of estimating the click timing.

\item $c_{\mu}$ represents the user's implicit aim point. If $c_{\mu}$ is 0 (or 1.0), it means that the user estimated on average the click timing the moment the pointer first contacted (or first exited) the target. A $c_{\mu}$ of 0.5 gives the lowest error rate, but users typically have a $c_{\mu}$ value lower than 0.5 \cite{lee2016modelling,lee2018moving,lee2019geometrically}, which is called the negative mean asynchrony (NMA) phenomenon \cite{vorberg1996modeling, repp2005sensorimotor, kim2018impact}.

\item $\nu$ is the rate at which the user encodes sensory information to estimate click timing from the \emph{visual cue}. The higher this value, the more reliable the user can perform with a shorter cue viewing time ($t_{c}$).

\item $\delta$ is the minimum standard deviation of the click timing estimated from the visual cue when the user is given enough cue viewing time ($t_c$).

\end{itemize}

\section{Implementation of Prediction Software}
This section describes how to predict the pointing error rate on a real system using the ICP model.
We need to implement a software that runs in three steps: (1) \emph{real-time trajectory logging}, (2) \emph{submovement segmentation}, and (3) \emph{predicting pointing error rate}.

\subsection{Step 1: Real-Time Trajectory Logging}
The software first logs the trajectory of the target and the pointer in real time until a click event occurs.
If the sampling rate of the input device being used is $f$ (unit: Hz), more specifically the software collects the following data from the moment of the previous click until the next click is observed:
\begin{equation*}
\centering
\left[ x_{p}^{i}, \, y_{p}^{i}, \, x_{t}^{i}, \, y_{t}^{i}, \, t^{i}\right], \, \text{for $i$ = $1$ to $N$}
\end{equation*}
($x_p^i, y_p^i$) and ($x_{t}^{i},y_{t}^{i}$) represent the coordinates of the $i$-th sampled pointer and the target, respectively. $t^{0}$ is the moment the preceding click occurred and $t^{N}$ is the moment when the current click event occurred. When a total of $N$ points are sampled, $t^{i}$ is the time stamp of the $i$-th sampling (i.e., $t^{i}=1/f \cdot i$).

\subsection{Step 2: Submovement Segmentation}
When a click occurs, the speed profile of the pointer is obtained by numerically differentiating the logged trajectory of the pointer.
The resulting speed profile is often contaminated with sensor noise from the input device. Since this reduces the performance of submovement segmentation, noise in the speed profile must be removed through a low-pass filter.
Various types of filters can be applied, but simple forms such as Gaussian kernel filters (e.g., $\sigma=3$) can work well enough.

\begin{figure}[t] 
	\centering
    \includegraphics[width=0.9\columnwidth]{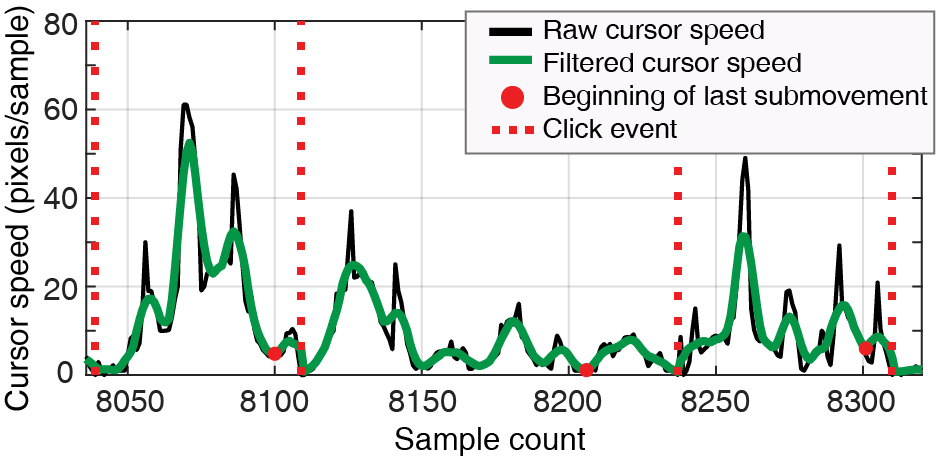}
    	\caption{The actual result of the submovement segmentation}
  ~\label{fig:submovement}
     ~\vspace{-5mm}
\end{figure}

The system then identifies the local minima and maxima in the smoothed speed profile and each neighboring minimum-maximum-minimum triplet is considered to be a possible submovement (see Figure \ref{fig:submovement}).
We use \texttt{Persistence1D} \cite{kozlov2015persistence1d} as an algorithm to find local extrema in the speed profile.
This algorithm returns all pairs of minima and maxima that exceed the pre-defined persistence value (e.g., 0.2 for a computer mouse).
To prevent jitter of click motion from being missegmented into a submovement, only triplets with a duration of at least 50 ms are considered as submovements.
\subsection{Step 3: Predicting Pointing Error Rate}
Among the submovements segmented in the previous step, the submovement started just before the click event is analyzed to predict the error rate (i.e., the last submovement).
If the last submovement of the pointer started at the $s$-th sample, the timestamp at that moment is $t^s$.
Because the user could observe the movement of the pointer during the last submovement, the cue viewing time ($t_{c}$) can be calculated from Equation \ref{eq:tc}:
\begin{equation}
t_{c}=t^{N}-t^{s}    
\end{equation}
The software then computes the average velocity vector of the pointer ($\vec{v}_{p}$) and target ($\vec{v}_{t}$) during the submovement:
\begin{equation*}
\begin{split}
&\vec{v}_{p}=\left[ (x_{p}^{N}-x_{p}^{s}),  \, (y_{p}^{N}-y_{p}^{s})\right]/t_{p}\\
&\vec{v}_{t}=\left[ (x_{t}^{N}-x_{t}^{s}),  \, (y_{t}^{N}-y_{t}^{s})\right]/t_{p}
\end{split}
\end{equation*}
The relative velocity vector $\vec{v}$ between the target and the pointer can be simply obtained as follows:
\begin{equation}
\vec{v}=\vec{v}_{p}-\vec{v}_{t}
\end{equation}
From the point ($x_{p}^{s}, y_{p}^{s}$) where the last submovement starts, the pointer approaches the target with a relative velocity of $\vec{v}$. The software then calculates $W_t$, which is the time it takes the pointer to pass through the target. $W_{t}$ can be obtained by dividing $W_{intersect}$ by the magnitude of the relative velocity $\lVert\vec{v}\rVert$, where $W_{intersect}$ is the length of the line segment defined by the intersection of the extended straight line $\vec{v}$ and the target (Figure \ref{fig:winterect}).
Next, $P$ is calculated, which is the period in which the click is repeated. This can be calculated as the average of the time intervals between all preceding clicks \cite{lee2016modelling, lee2018moving}.

From the obtained $P$, $t_c$, and $W_t$, the software can finally predict the pointing error rate through Equation \ref{eq:er}. If free parameters are already available, the error rate is calculated immediately. If free parameters are not obtained in advance, the values can be obtained through a general optimization process that minimizes the difference between the error rate predicted by the ICP model and the error rate of users observed in actual experiments. In the following sections, we demonstrate those processes in greater detail by conducting actual user studies and verify the error rate prediction performance of the model.

\section{Study 1: Pointing on a Stationary Target}
\begin{figure}[!t] 
	\centering
    \includegraphics[width=1.0\columnwidth]{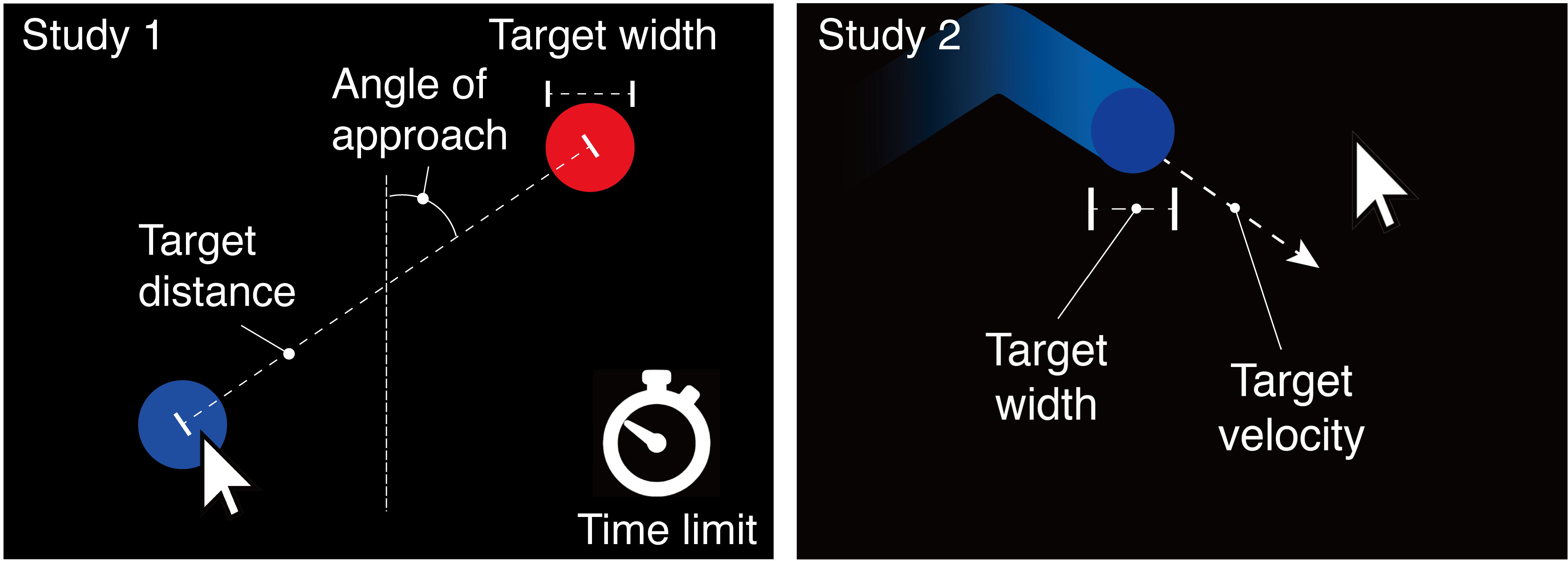}
    	\caption{Task screen of Study 1 (pointing on a stationary target) and Study 2 (pointing on a moving target)
}
  ~\label{fig:study1screen}
   ~\vspace{-5mm}
\end{figure}

Based on the implemented system, two user studies were conducted (Figure \ref{fig:study1screen}).
In all studies, the data analysis was performed using the implementation described in the previous section.
In Study 1, participants performed a two-dimensional pointing task on a stationary target.
Users were given a time limit, which resulted in a wide range of error rates.
We used Wobbrock's error rate model \cite{wobbrock2008error} as the baseline for performance comparison.

\subsection{Method}
\subsubsection{Participants}
Twelve paid participants from a local university (7 males, 5 females) were recruited. 
The average age of the participants was 24.42 years ($\sigma$=3.26).
All of the participants were right-handed.
Their average mouse usage time per day was 5.63 hours ($\sigma$=3.59).
The participants played games 6.86 hours ($\sigma$=2.61) per a week with a computer mouse.
All participants had normal or corrected vision.

\subsubsection{Design}
The experiment followed a 2$\times$3$\times$6 within-subject design with three independent variables: \emph{Target Width}, \emph{Target Distance}, and \emph{Time Limit}.
The levels were the following:
\begin{itemize}
\item Target Width: 4.8 and 8.4 mm
\item Target Distance: 48, 144, and 240 mm
\item Time Limit: 300, 400, 500, 600, 700, and 800 ms
\end{itemize}

Twenty angle of approaches were tested for each \emph{Time Limit}-\emph{Target Width}-\emph{Target Distance} condition.
A \emph{Time Limit} condition did not change to the next condition until all corresponding width-distance conditions had been completed (240 pointing trials per each \emph{Time Limit} condition).
Within a \emph{Time Limit} condition, the \emph{Target Width} and \emph{Target Distance} are given in random order.
The \emph{Time Limit} conditions are given in a random order.
Each \emph{Time Limit} condition was repeated twice.
The angle of approach was given to the participant in a clockwise sequence of 360 degrees divided into 20 steps.
In the end, 17,280 (=$2\times3\times6\times2\times20\times12$) input events from 12 participants were logged.

\subsubsection{Task} 
Participants were given two circular targets.
After clicking on the blue target, clicking on the red target ended the trial (Figure \ref{fig:study1screen}).
If the participant did not click on the red target within the given \emph{Time Limit} after clicking on the blue target, the red target disappeared.
Even if the red target disappeared, participants had to click to go to the next trial.
If a participant clicked inside the red target (or the disappeared red target), the trial was considered successful.
Participants were asked to perform pointing as quickly and accurately as possible.
They were also asked to complete each trial within the \emph{Time Limit}.

\subsubsection{Apparatus}
The application was implemented on a 3 GHz desktop computer (\texttt{Mac mini}, 10.13.1).
A 27-inch LED monitor (\texttt{LG 27UD69P}) was used, and the resolution of the task screen was 2,560$\times$1,440 pixels.
The pointing device was a two-button wired optical mouse (\texttt{Samsung SNJ-B138}) with a resolution of 1,000 DPI and a the polling rate of 125 Hz.
The pointer was a standard arrowhead pointing to the upper left.
The mouse gain function maintained the default setting of the OS.
The refresh rate of the application was maintained at 60 Hz.

\subsubsection{Procedure}

Participants sat on a regular office chair and the monitor was installed at the participant's eye level.
Before the experiment, the experimenter briefly introduced the task to the participants.
Subsequently, the participants filled out a pre-questionnaire.
Participants also signed a consent form.
A practice session was given until participants were accustomed to the task.
The experiment for each individual took about an hour and each participant was rewarded with \$10.

\subsection{Results}
\subsubsection{Data Validation}
For all trials, the movement time that the participants actually performed the task was about 124\% of the given \emph{Time Limit}.
However, as the last submovement already started at 72\% (SD=51\%) of the \emph{Time Limit}, we know that the participants did not intentionally wait for the target to disappear.
\finalDel{52 trials were considered accidental clicks and were removed (0.3\% of the total data).
No other data was removed.}

\subsubsection{Descriptive Statistics}
The overall average error rate for all participants' trials was 37\%.
This is about two times higher than the error rate in other studies \cite{wobbrock2008error, wobbrock2011modeling}.
The duration of the last submovement, or the cue viewing time $t_{c}$, was measured to be 296 ms (SD=119 ms) on average, which is similar to the known values in previous studies \cite{jagacinski1980fitts, lee2016autogain}.
$W_{t}$ was measured to be 244 ms on average (SD=285).
In 2,613 trials (15.1 \%), the pointer moved in a direction that could not intersect the target (i.e., $W_{intersect}$=0).
In that case $W_t$ was considered zero.
Except for those cases, the average of the measured $W_t$ was 288 ms (SD=289).
The mean of the input repetition period $P$ was measured as 636 ms (SD=207).
The average magnitude of the relative velocity between the target and the pointer was 64.98 mm/s (SD=103).

\subsubsection{Overall Model Fit to the Baseline Model}
As a baseline, Wobbrock's model \cite{wobbrock2008error,wobbrock2011modeling} was fitted to 36 data points using Equation \ref{eq:wobbrockmodel} (2 \emph{Target Widths} $\times$ 3 \emph{Target Distances} $\times$ 6 \emph{Time Limits}).
In the equation, $MT_e$ is calculated as the mean time actually spent in a pointing trial at each condition, not the given \emph{Time Limit} value.
We used the \texttt{Global Optimization Toolbox} of \texttt{MATLAB} for the fitting.

As reported in the previous studies \cite{wobbrock2008error,wobbrock2011modeling}, the model successfully explains the error rate of the user (\finalDel{$R^{2}$ = 0.955,}\finalAdd{Adjusted $R^{2}$ = 0.954,} see Figure \ref{fig:study1result}).
The free parameters of the model were $a$=130 ms and $b$=157 ms/bit, respectively.
Because these values are based on Fitts' law, throughput can be calculated as the reciprocal of $b$.
As a result, we obtained 6.37 bit/s similar to the previously measured value for the mouse \cite{mackenzie2008fitts}.
\begin{table}[!b]
\centering

\caption{The fitting results from the two user studies in this paper
}
\label{table}

\begin{tabular}{|c|c|c|c|c|c|}
\hline
                                    & $c_\mu$   & $c_\sigma$ & $\nu$     & $\delta$  & $R^{2}$ \\ \hline

Study 1                                     &    0.129     &  0.0873   &    14.532    &  0.461     & 0.992
              \\ \hline              
              Study 2                                     &    0.241     &  0.093  &    25.33    &  0.337     & 0.985
              \\ \hline

\end{tabular}
\end{table}
\subsubsection{Overall Model Fit to ICP Model}
In our model, the variable that determines the error rate of the user is $W_{t}/D_{t}$ (see Equation \ref{eq:er}).
Therefore, the total data is arranged in order of $W_{t}/D_{t}$, and then the error rate is obtained through the sequential binning from the left.
This allows us to get 36 final data points as we fit the baseline model.
The following values from a previous study \cite{lee2018moving} were used as the free parameters for the initial sorting: $c_{\mu}=$0.25, $c_{\sigma}=$0.08, $\nu=$20.2, and $\delta=$0.366.

\begin{figure}[!t] 
	\centering
    \includegraphics[width=1.0\columnwidth]{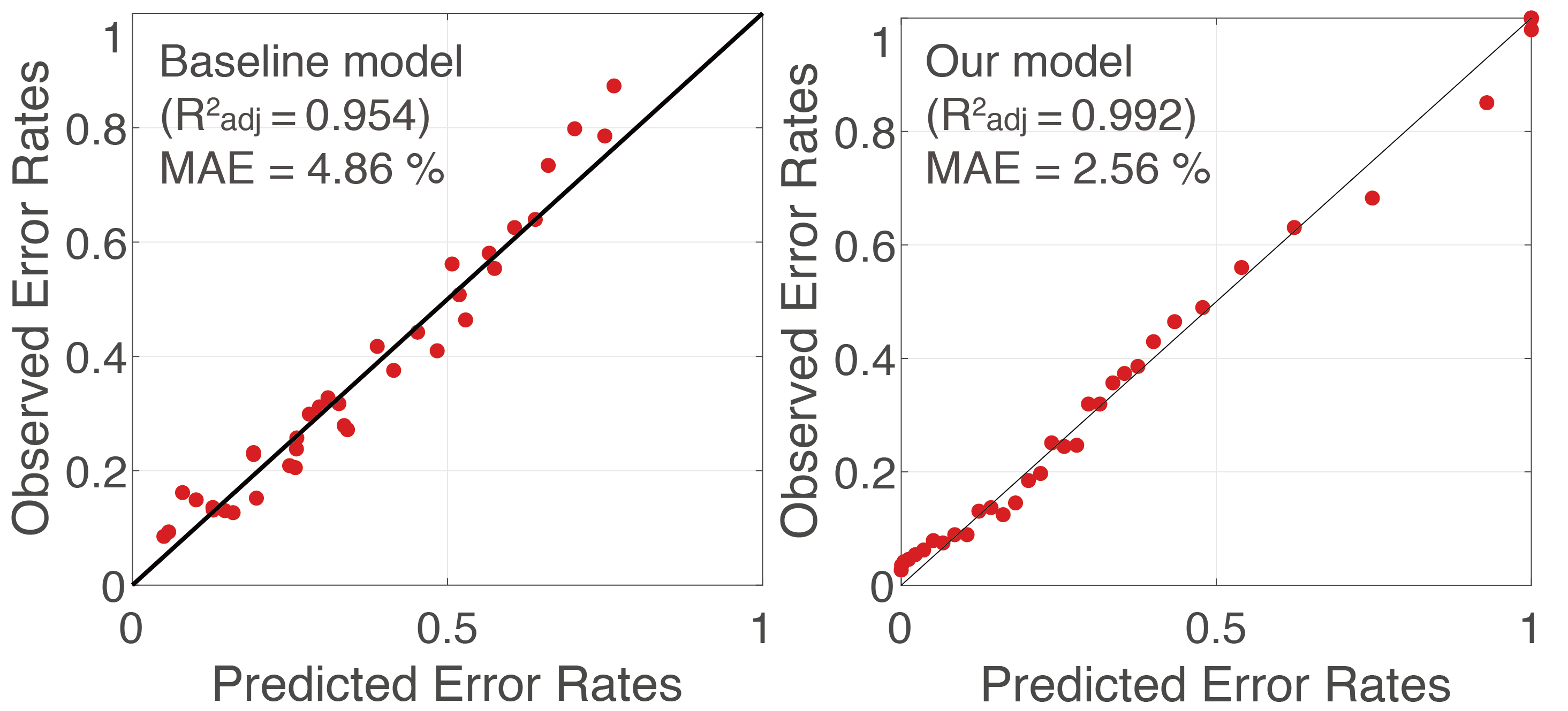}
    	\caption{Results from Study 1: Both the baseline model and the model proposed in this study explained the observed error rate well. 
}
  ~\label{fig:study1result}
   \vspace{-5mm}
\end{figure}

As a result, our model fitted with the observed error rate with a high coefficient of determination (\finalDel{$R^{2}=0.993$,}\finalAdd{adjusted $R^{2}=0.992$,} see Figure \ref{fig:study1result}).
The free parameters obtained as a result of fitting are summarized in the Table \ref{table}.
We also conducted two-fold cross validation with random sampling for each model.
The mean absolute error (MAE) values obtained were 4.86\% and 2.56\% for the baseline model and our model, respectively.

\subsection{Discussion}
In the task of pointing to a stationary target, the baseline model and the ICP model predicted the error rate of the participants with high accuracy.
However, our model predicted error rates more accurately than the baseline model.
This supports our hypothesis that the pointing error occurs from the user's click timing estimation during the last submovement.

\section{study 2: Pointing on a Moving Target}
In Study 2, participants pointed to a circular moving target.
Regardless of the target movement, we analyzed the last submovement with the same algorithm as in Study 1 and fitted it to the model.
We used Huang's model \cite{huang2019modeling} as a baseline.

To verify that the free parameters obtained from the fittings successfully reflect the cognitive characteristics of the users, we recruited participants into two groups: gamers and non-gamers. Participants in the gamer group were recruited as experts in the first-person shooter (FPS) games due the similarities of the tasks performed in such games and in the study.

\subsection{Method}
\subsubsection{Participants}
We recruited a total of 16 participants divided into two groups: (1) gamers (8 males) and (2) non-gamers (1 male, 7 females). 
All the participants were right-handed.
The average age of participants in the gamer group was 24.4 years (SD=3.81) and in the non-gamer group was 25.63 years (SD=4.53). 
Participants of the gamer group played FPS games an average of 15 hours per week, and their average mouse usage time per day was 7.13 hours (SD=2.23). Meanwhile, non-gamer group participants did not play FPS game at all and they use a mouse for an average of 4.25 hours (SD=3.99) per week.

\subsubsection{Participant Recruiting Criteria}

We recruited gamers based on the following criteria: (1) a player of the game \texttt{PlayerUnknown's Battlegrounds} who was within the top 5\% rating, or (2) a player of the game \texttt{Overwatch} who owned a higher level than master (the top 2 to 3\%) and who mainly focused on characters who need excellent aiming ability (such as Hanzo or Mccree).
Meanwhile, non-gamers were recruited with people who had never played FPS games before.

\subsubsection{Design}
The experiment followed a within-subject design with two independent variables: \emph{Target Speed} and \emph{Target Width}.
These two factors were randomly determined for each trial in the following ranges:
\begin{itemize}
\item Target Speed: from 0 mm/s to 510 mm/s
\item Target Width: from 9.6 mm to 24 mm 
\end{itemize}
In order to satisfy ecological validity, we reproduced the speed range of the target in commercial games such as \texttt{Fruit Ninja} (107 mm/s) and \texttt{Piano Tiles}  (314 mm/s).
The location where the target was generated and the direction the target moved were randomly determined for each trial.
Participants performed a total of 9 \emph{Block}s of trials and each \emph{Block} consisted of 200 trials.
As a result, 28,800 input events from 16 participants were logged (=$16 \times 9 \times 200$). 
\subsubsection{Task}
Participants were instructed to click on a blue circular target moving straight at a constant speed (see Figure \ref{fig:study1screen}).
If the target collided with a wall (edge of the screen), the target bounced at the same angle as the incident angle.
The trial was considered successful only when the participant clicked inside the target.
Regardless of success, if a click event occurred, the current target disappeared and a new target was created with randomized size and speed.
Participants were asked to perform pointing as quickly and accurately as possible.        

\subsubsection{Apparatus}
The application was implemented on the same 3 GHz desktop computer as in Study 1 (\texttt{Mac mini}, 10.13.1 High Sierra). 
A 24-inch LED-backlit LCD gaming monitor (\texttt{ASUS ROG SWIFT PH248Q}) was used and the resolution of the task screen was 1920$\times$1080 pixels.
The pointing device was a wired optical mouse (\texttt{Logitec G502}) with a resolution of 1000 DPI and a polling rate of 125 Hz. 
Mouse acceleration was disabled and the tracking speed of the mouse was 4/10 (the default setting of the \texttt{Mac OS}).
The refresh rate of the application used in the experiment was maintained at 60 Hz.

\subsubsection{Procedure}
Participants sat on a regular chair.
The display was installed at the participant's eye level.
Participants signed a consent form before the experiment.
After the participants completed the pre-questionnaire, the experimenter briefly demonstrated the task.
50 trials were given to participants as a practice session before starting the main study.
A one-minute break was provided at the end of each \emph{Block}.
It took about an hour per person to finish the study.
The amount of compensation for participation was the same as in Study 1.

\subsection{Results}
\finalDel{\subsubsection{Data Validation}
1,032 trials with trajectory were considered accidental clicks and were removed (3.58\% of the total data).
Due to the dynamic nature of the task, the ratio of accidental clicks was higher than Study 1.
No other data has been removed.}

\begin{figure}[!t] 
	\centering
    \includegraphics[width=1\columnwidth]{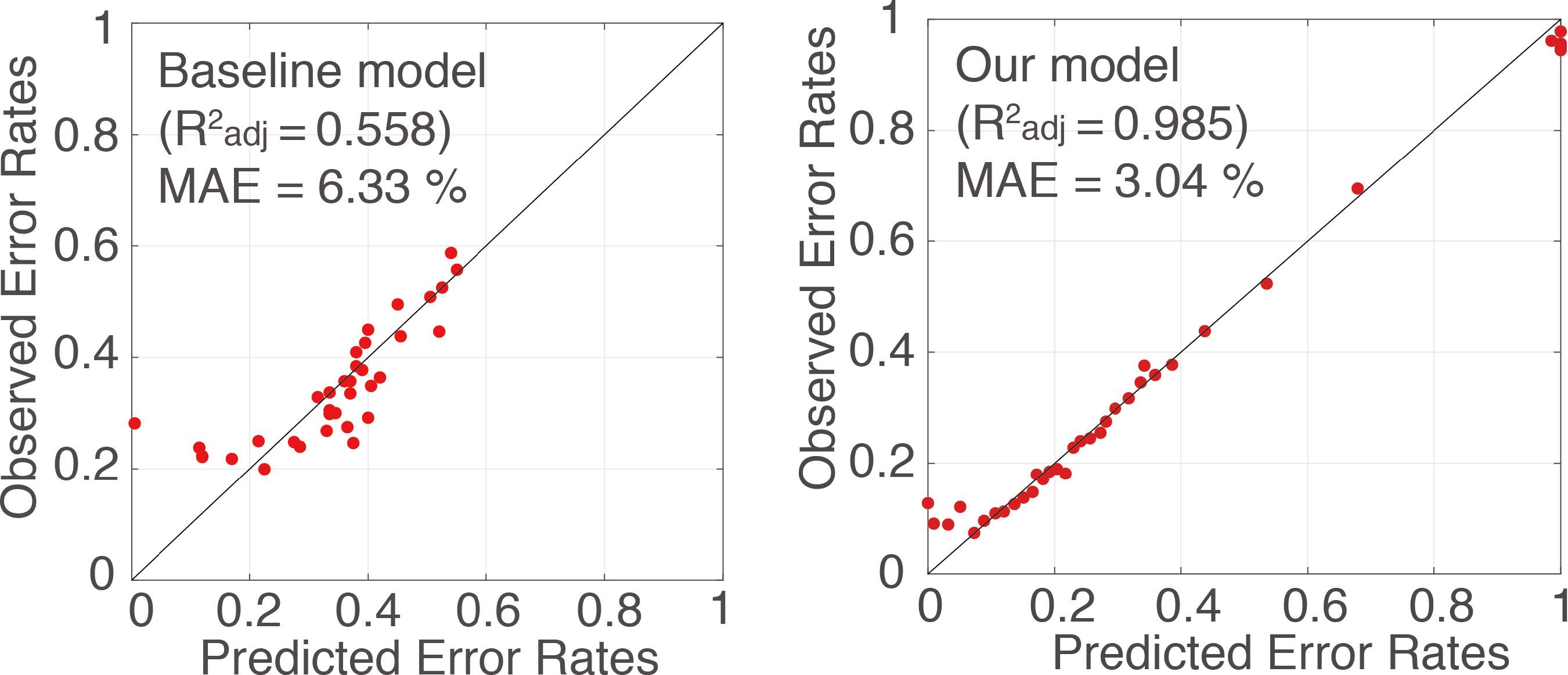}
    	\caption{Results from Study 2}
  ~\label{fig:study2result}
   ~\vspace{-7mm}
\end{figure}

\subsubsection{Descriptive Statistics}
The overall average error rate for all participants' trials was 37\%.
This value was almost the same as in Study 1.
The duration of the last submovement, or $t_{c}$, was measured to be 275 ms (SD=238 ms) on average, which is similar to the submovement duration reported in previous studies \cite{jagacinski1980fitts, lee2016autogain}.
$W_t$ was measured to be 107 ms on average (SD=148 ms).
In 4,658 trials (16.2 \%), the pointer moved in a direction that could not intersect the target (i.e., $W_{intersect}$=0).
In that case, $W_t$ was considered zero.
The average interval from click to click was 905 ms (SD=460 ms).
The average magnitude of the relative velocity between the target and the pointer ($\lVert\vec{v}\rVert$) was 144 mm/s (SD=74 mm/s).

\subsubsection{Removing Learning Effect}
Pointing to a moving target is a challenging task, so there can be a significant learning effect.
In fact, the effect of \emph{Block} on error rate was significant (F(8,120)=5.183, p<0.001).
From the Helmert contrast of the \emph{Block} effect, we confirmed that the learning effect becomes insignificant from the third \emph{Block} (p=0.376)
The following results were obtained by analyzing the remaining seven \emph{Block}s after excluding the first two.

\subsubsection{Overall Model Fit to the Baseline Model}
We used Huang's model as the baseline model for the Study 2 \cite{huang2019modeling}.
In Huang's model, the difficulty of a task can be determined according to $\mu_t$, $\sigma_t$, and $\sigma_n$ (see Related work section). Therefore, we arranged the data by the average of these three values. Then, we binned in order from the smallest value to the larger one and obtained 36 fit points to plot. In this study as well, we used the \texttt{Global Optimization Toolbox} of \texttt{MATLAB} for the fitting. Huang's model does not provide explicit error rate prediction for circular targets. Therefore, we calculated the error rate based on Monte-Carlo simulation from the distribution predicted by the model and the average target size of the bin.
The data was fitted to the baseline model with \finalAdd{adjusted $R^{2}$=0.558} and the MAE value obtained from two-fold cross validation was 6.33\%. The free parameters of the model were $a_{t}$=0.13, $b_{t}$=1.0$^{-4}$, $c_{t}$=-0.19, $d_{t}$=3.6$^{-10}$, $e_{t}$=0, $f_{t}$=0.041, $g_{t}$=0, $d_{n}$=0.003, $e_{n}$=0, and $f_{n}$=0. 

\subsubsection{Overall Model Fit to the ICP Model}
In the same manner as in Study 1, the data of all trials were binned by $W_{t}/D_{t}$, and finally, 36 averaged data points were obtained.
As a result, our model fitted the empirical error rate with a high coefficient of determination (\finalAdd{adjusted $R^{2}=0.985$,} see Figure \ref{fig:study2result}).
The free parameters obtained as a result of fitting are summarized in the Table \ref{table}.
The free parameters obtained were similar to those in the previous study \cite{lee2018moving} and in Study 1.
We performed two-fold cross validation with random sampling.
The MAE was 3.04\% for our model.

\subsubsection{Comparing Gamers and Non-gamers}
\begin{figure}[!t] 
	\centering
    \includegraphics[width=1\columnwidth]{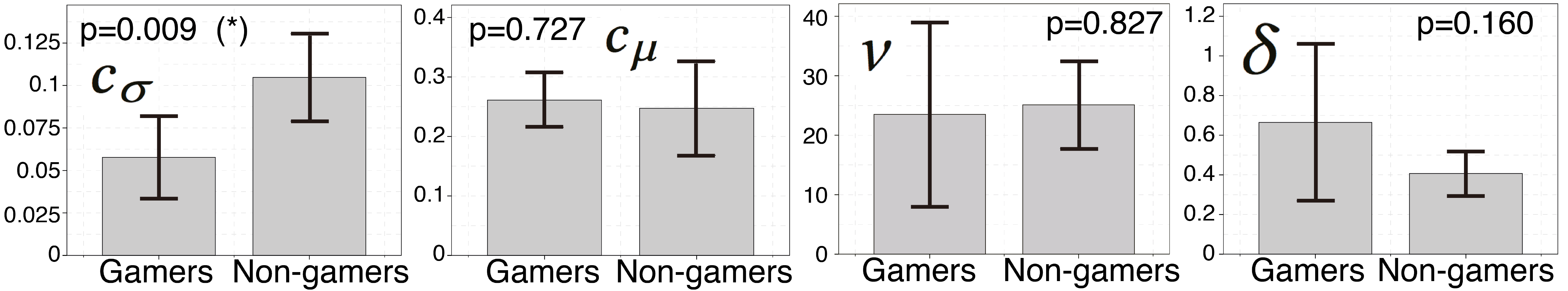}
    	\caption{Comparison of four free parameters between the gamer group and the non-gamer group}
  ~\label{fig:compare_gamers}
  ~\vspace{-4mm}
\end{figure}

By fitting our model to individual data, we obtained four free parameters ($c_\sigma,c_\mu,\nu,\delta$) for each participant.
Considering the reduced number of individual data points, the error rates were obtained from the 18 bins of $W_{t}/D_{t}$.
As a result, the error rate for each participant was fitted to the model with a high coefficient of determination ($R^{2}$=0.938 to 0.990, M=0.973, SD=0.016).
An independent-samples t-test was conducted to compare free parameters and error rates between gamers and non-gamers.
There was a significant difference in the error rates for gamer (M=27.4\%, SD=6.9\%) and non-gamer (M=44.1\%, SD=14.1\%) groups; $t$(14)=3.021, $p$ = 0.009.
There was no significant difference in the period of input repetition ($P$) between gamers (M=871 ms, SD=186 ms) and non-gamers (M=886 ms, SD=64 ms); $t$(8.641)=-0.218, $p$ = 0.833.
There was also a significant difference in the $c_\sigma$ for gamer (M=0.058, SD=0.029) and non-gamer (M=0.11, SD=0.031) groups; $t$(14)=-3.14, $p$ = 0.007.
No significant difference was found for the other three parameters $c_\mu$ ($t$(14)=0.36, $p$=0.73), $\nu$ ($t$(14)=-0.22, $p$=0.83), and $\delta$ ($t$(14)=1.48, $p$=0.16).
The results are summarized in Figure \ref{fig:compare_gamers}.

\subsection{Discussion}
Unlike Study 1, in Study 2, there was a significant difference in explanatory power between the ICP model and the baseline model. The results of the cross validation also showed that the ICP model predicted the error rate better than the baseline model. The baseline model, however, predicted the error rate in moving target pointing with high accuracy in the original paper \cite{huang2019modeling} ($R^{2}$=0.94). The difference between their results and ours seems to be due to differences in the speed range of the targets used in the experiment. Their experiment used targets with a maximum velocity of approximately 200 mm/s when the pixels were roughly converted to millimeters. However, our experiments used a target speed of up to 500 mm/s, which is 2.5 times faster than the target speed they used.

Huang's model predicts that the faster the target speed ($V$ in Equation \ref{eq:huangmodel}), the greater the variance of the pointing end point. However, according to the ICP model, even if the target moves fast, the pointing error rate can be lowered if the user moves well to reduce the relative speed between the pointer and the target. In fact, the coefficients of the Huang model, which are multiplied by the target velocity ($e_{t}$, $e_{n}$, $g_{t}$), were mostly optimized to be zero at the fitting, indicating that the variance of the endpoint distribution did not increase as the model predicted when the target velocity was increased.

The parameters of the ICP model obtained from Study 1 were very similar to the values obtained from Study 2 (see Table \ref{table}).
This supports the assumption of the ICP model that the planning and execution of click actions is performed with the same cognitive-motor process regardless of whether the target is moving or not.

The model also succeeded in uncovering differences in cognitive characteristics in the click process between gamers and non-gamers.
\finalAdd{Compared to non-gamers, gamers had higher $c_{\sigma}$ value and lower $\delta$ value. As shown in Equation \ref{eq:variances}, the reliability of the click timing obtained from the visual cue ($\sigma_{v}$) can be thought of as the product of $c_{\sigma}$ and $\delta$. However, the reliability of the click timing obtained from the temporal cue ($\sigma_{t}$) is determined from $c_{\sigma}$ alone. As shown in Figure \ref{fig:compare_gamers}, multiplying the $c_{\sigma}$ and $\delta$ values yields similar values for gamers and non-gamers. Therefore, we can interpret that gamers have similar ability to estimate click timing from visual cues to that of non-gamers, but have better ability to encode the rhythm of clicks with an internal clock.}

\finalDel{The fact that gamers have a lower $c_\sigma$ value than non-gamers shows that even if the same sensory signal is given, the internal clock of the gamer can more precisely estimate optimal click timing than non-gamers. This shows why the expert is superior to the novice in FPS performance \cite{Vicencio-Moreira:2014:EAT:2556288.2557308}.}

\vspace{-1mm}
\section{Conclusion}
The model proposed in this study accurately predicted users' pointing error rate with a simple algorithm regardless of the target motion ($R^{2}$= 0.985 to 0.992 and MAE=2.56\% to 3.04\%).
In particular, the four free parameters obtained from the data fitting remained similar for different pointing situations (see Table \ref{table}).
Based on this robust explanatory power, the model revealed significant cognitive differences between gamers and non-gamers.

Nonetheless, this study has some limitations.
First, it is difficult to apply our model in a situation where the trajectory of the cursor is difficult to track.
Secondly, further validation is needed as to whether this model is generally applicable for more complex patterns of target motion.
Third, our model did not explain how a user's internal clock encodes the temporal structure cue when the input is randomly repeated.
\finalAdd{Fourth, the ICP model was currently only tested for young users in their 20s. For users from other age groups, further research will be needed to determine whether the ICP model predicts error rates well and extracts meaningful parameters. Fifth, the ICP model only predicts the error rate of the click action and does not predict how the user's tracking movement will be performed before the click action. We envision that this limitation can be overcome by integrating the ICP model with existing control theoretical \cite{muller2017control, bye2008bump, bye2010bump} pointing models that can simulate the user's target tracking movements.}

\finalAdd{
\section{Acknowledgements}
This research was funded by the National Research Foundation of Korea (2017R1C1B2002101) and the Korea Creative Content Agency (R2019020010).}

\balance{}
\bibliographystyle{SIGCHI-Reference-Format}
\bibliography{sample}

\end{document}